\documentclass[11pt]{article}
\usepackage{moriond,epsfig}
\usepackage{amsmath}
\usepackage{amssymb}

\def\sst{\scriptscriptstyle}

\def\be{\begin{equation}}
\def\ee{\end{equation}}
\def\bea{\begin{eqnarray}}
\def\eea{\end{eqnarray}}

\newcommand{\mpl}{M_{\rm\scriptscriptstyle Pl}}

\begin{document}
\vspace*{4cm}
\title{NEUTRALINO RELIC DENSITY ENHANCEMENT\\ IN NON-STANDARD COSMOLOGIES~\footnote{Talk presented by S.~Profumo at the XXXIXth Rencontres de Moriond on {\em Exploring the Universe}, La Thuile, Italy, March 28 - April 4, 2004.}}

\author{ S.~PROFUMO and P.~ULLIO }

\address{SISSA/ISAS, via Beirut 2-4, I-34013 Trieste, Italy\\ and INFN, Sezione di Trieste, I-34014 Trieste, Italy}

\maketitle\abstracts{The thermal relic abundance of species critically depends on the assumed underlying cosmological model. In the case of neutralinos, freeze-out takes place long before Big Bang Nucleosynthesis, which provides the strongest constraint on the evolution of the Hubble parameter in the Early Universe. We show that non-standard cosmologies, such as models featuring a quintessential scalar field or primordial anisotropies, can lead to large enhancements in the neutralino relic abundance, up to six orders of magnitudes. Within these scenarios, supersymmetric models with large neutralino annihilation cross sections may account for the whole inferred amount of cold dark matter, yielding on the other hand large indirect detection rates.}

\section{Introduction}

Observational cosmology is accumulating evidences for a picture of the Universe dominated by a {\em dark sector} made up of a non-luminous and non-baryonic component ({\em dark matter}) and by a gravitationally non-clustering form of energy with negative pressure ({\em dark energy})~\cite{Spergel:2003cb}. While the standard cosmological lore, often dubbed {\em Concordance} or $\Lambda$CDM {\em Model} nicely fits most of the observational features~\cite{giavalisco}, particle physics and theoretical cosmology face the compelling questions related to the nature and the amount of both dark matter and dark energy. 

The minimal supersymmetric extension of the Standard Model provides a suitable candidate for cold dark matter, namely the lightest neutralino. The thermal relic abundance of neutralinos $\Omega_\chi h^2$ depends both on the particle physics setup, and therefore on the supersymmetric (SUSY) model, and on the underlying cosmological model, which affects the freeze-out of species in the Early Universe. Models where the r\^ole of dark energy is played by a dynamical scalar field ({\em Quintessence}), or scenarios where extra energy density components are present at the time of neutralino freeze-out, may yield significant modifications in the relic abundance of neutralinos~\cite{Profumo:2003hq}. This in turns translates in the intriguing possibility that SUSY models with large neutralino annihilation cross sections, and hence with copious dark matter detection rates, give rise to the required amount of cold dark matter. 

\section{Cosmological Enhancement of Neutralino Relic Abundance}

\begin{figure}
\begin{center}
\mbox{\hspace{-.5cm}\epsfig{file=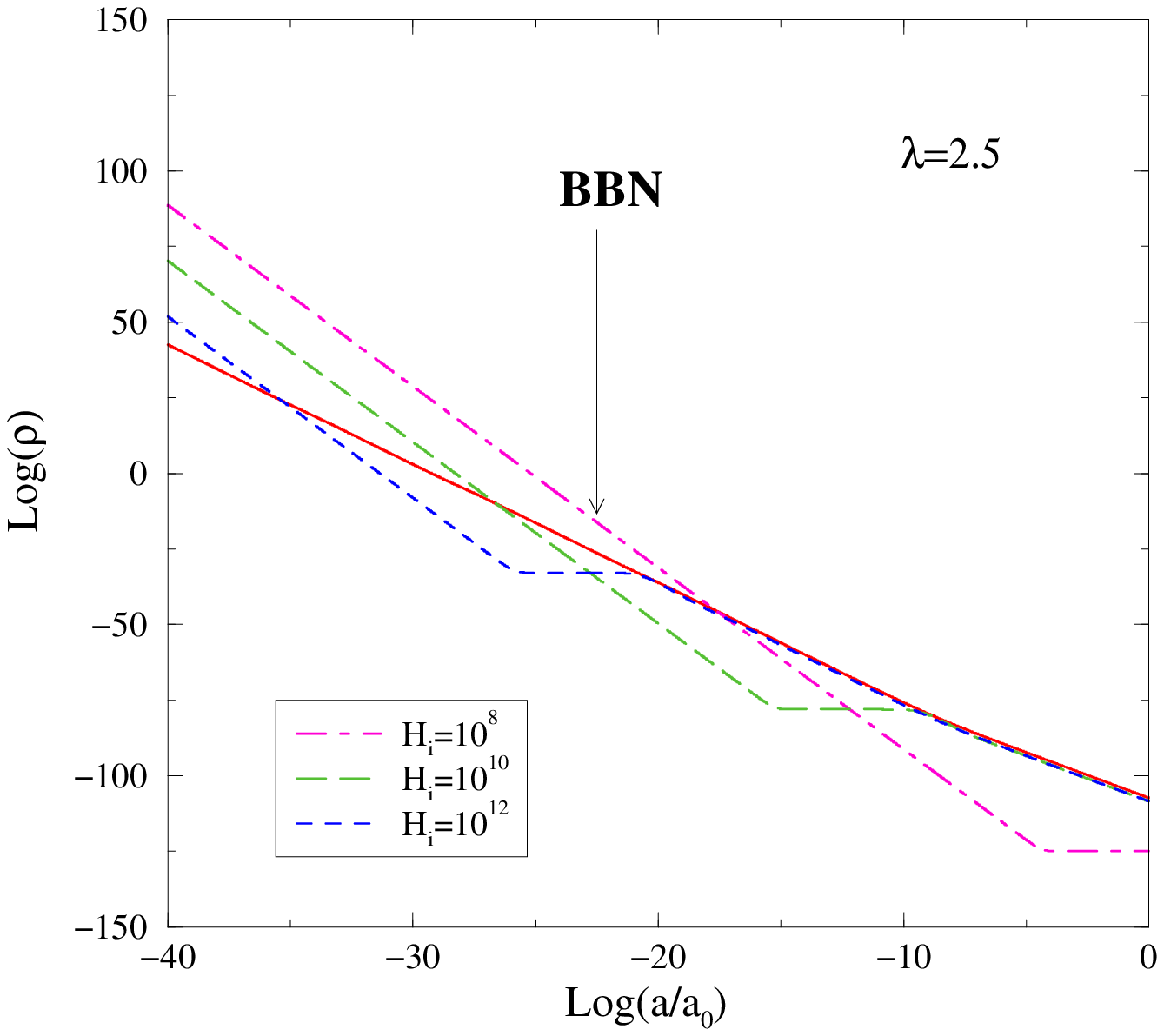,width=6.5cm}\hspace{0.7cm} \epsfig{file=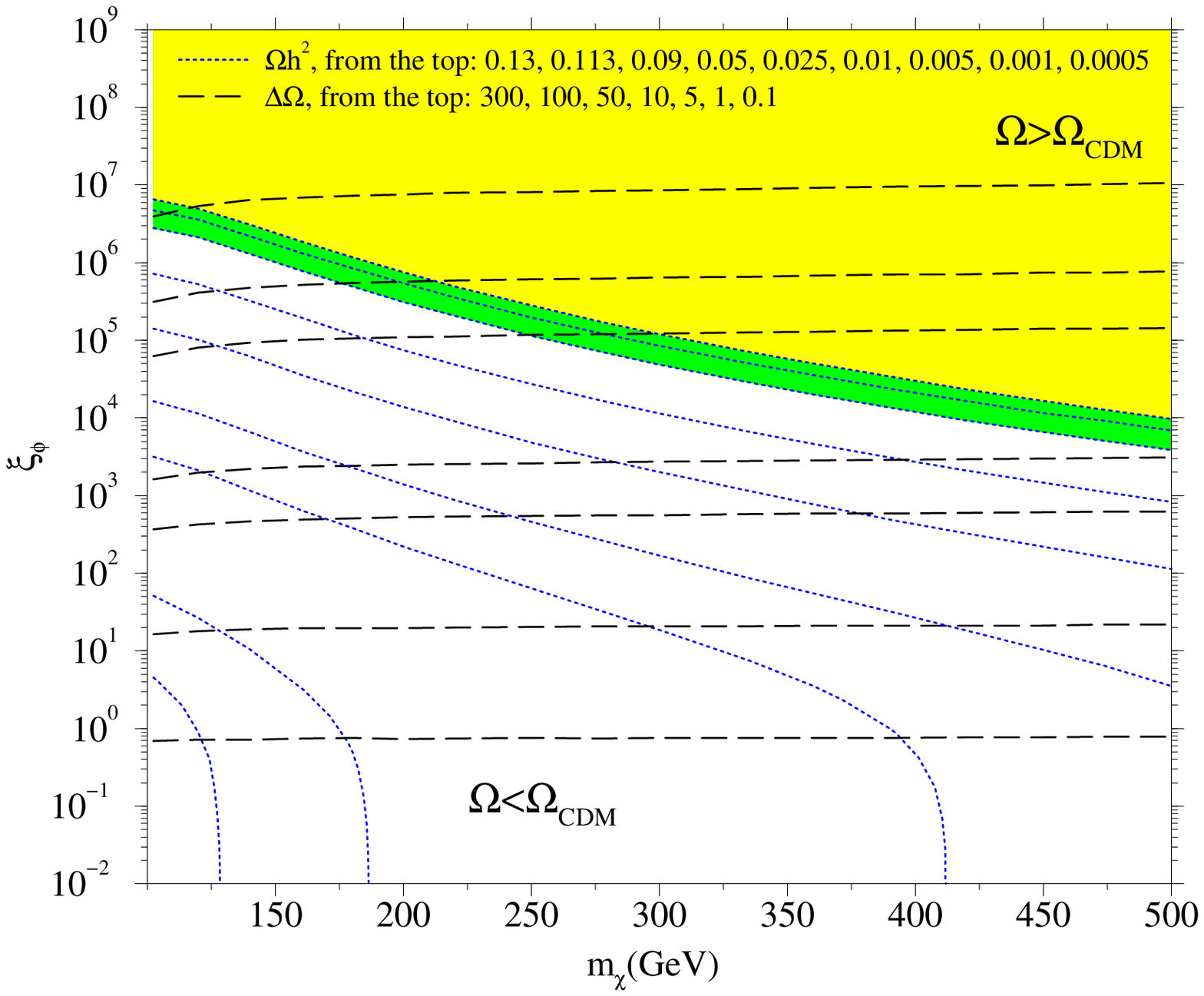,width=10cm}}\\
{\large (a)\hspace{8.5cm}(b)\hspace*{1.cm}}
\end{center}
\caption{(a): Sample cases of Cosmologies with a Quintessence field. The solid lines refer to the sum of the radiation and matter energy density components, the other curves to the Quintessence contributions at $\lambda=2.5$ and at three values of $H_i$. (b): Neutralino relic density enhancement in the minimal Anomaly Mediated SUSY Breaking Model, at $\tan\beta=50$, $\mu>0$ and $m_0=1\ {\rm TeV}$ and $m_{3/2}$ ranging from 32 to 192 TeV, as a function of the neutralino mass $m_\chi$ and of the parameter $\xi_\phi$ in the quintessential scenario with $\lambda=3.5$. The yellow shaded region has $\Omega_\chi h^2>0.13$, while in the green-shaded strip $0.09<\Omega_\chi h^2<0.13$. Blue dotted lines correspond to points at fixed $\Omega h^2$ while black dashed lines to points at fixed enhancements $\Delta\Omega$.
\label{fig1}}
\end{figure}

The thermal relic abundance of neutralinos is determined by a Boltzmann equation, in the form:
\begin{equation} \label{eq:Boltzmann}
  \frac{dn}{dt} =
  -3Hn - \langle \sigma_{\rm{eff}} v \rangle 
  \left( n^2 - n_{\rm{eq}}^2 \right).
\end{equation}
When the effective annihilation rate of the particles falls below the expansion rate of the Universe, 
\begin{equation} \label{eq:fo}
n_{\rm eq}\langle \sigma_{\rm{eff}} v \rangle(T) \lesssim H(T)
\end{equation}
the species ceases to be in thermal equilibrium and {\em freezes-out}, locking, after a residual annihilations transient, its comoving number density, and therefore its relic abundance today. While the left hand side (l.h.s.) of eq.~(\ref{eq:fo}) depends on the details of the particle physics model, the r.h.s. is fixed by the cosmological model. The Hubble rate depends in its turn, through the Friedmann equation (assuming a spatially flat Universe), on the energy density of radiation, matter and of any other component $X$ as
\begin{equation} \label{eq:fried}
   H^2\left(T\right) = \frac{1}{3 \mpl^2} 
   \left[\rho_{r}\left(T\right) + \rho_m\left(T\right) + 
   \rho_{\sst X}\left(T\right)\right]\;,
\end{equation}
In the presence of a single spatially-homogeneous scalar field $\phi$ with a potential 
$V\left(\phi\right)$, the equation of motion of the field is {\em dynamically coupled} to eq.~(\ref{eq:fried}), and 
\begin{equation} \label{eq:quint}
\rho_{\sst X}\rightarrow \rho_\phi\equiv\frac{1}{2} \left(\frac{d\phi}{dt}\right)^2 + 
   V\left(\phi\right).
\end{equation}
In the present study, we take as a representative case the exponential potential~\cite{Ferreira:1997hj}
\begin{equation}
   V\left(\phi\right) = \mpl^4 \exp\left(-\lambda \phi/\mpl\right)\;.
   \label{eq:exppot}
\end{equation}
The model then depends only on two parameters, namely the exponent $\lambda$ and the Hubble rate at the end of inflation, $H_i$, which fixes the initial ratio of the energy densities of the quintessential field and of radiation~\cite{Ferreira:1997hj}. The behavior of $\rho_\phi(a)$ is sketched in fig.~\ref{fig1} (a) for a few sample cases at $H_i\gg 1$: after a period of {\em kination}, where $\rho_\phi\sim a^{-6}$, the scalar field undergoes a {\em cosmological constant-like} behavior, and then possibly {\em tracks} the dominant energy density component. For a suitable choice of ($\lambda,H_i$), $\rho_\phi$ may be compatible with Big Bang Nucleosynthesis (BBN) constraints and provide the correct amount of dark energy at the present epoch~\cite{Profumo:2003hq,Ferreira:1997hj}, while giving a sizeable contribution to the energy density budget at neutralino freeze-out ($T_{\rm f.o.}\sim m_\chi/20\sim \mathcal O (10)$ GeV). 

To quantify the enhancement, it is convenient to introduce the parameter $\xi_{\sst X}$, defined as the ratio $\xi_{\sst X}\equiv\left(\rho_{\sst X}/\rho_r\right)(T^{{\rm No }X}_{\rm f.o.})$ of the extra-components $X$ and radiation energy densities at the freeze-out temperature of the neutralino, for definiteness taken to be that in the absence of extra-components. Fig.~\ref{fig1} (b) shows the neutralino relic abundance on a parameter space slice of the minimal Anomaly Mediated SUSY Breaking (mAMSB) scenario in the $(m_\chi,\xi_\phi)$ plane. In mAMSB, where the lightest neutralino $\chi$ is wino-like, the occurrence of a relic density enhancement mechanism is clearly mandatory to allow $\chi$ to be the main CDM constituent: fig.~\ref{fig1} (b) shows that, in presence of a dynamical quintessential scalar field, $\Omega_\chi h^2$ may well lie within the preferred WMAP range~\cite{Spergel:2003cb} for any $m_\chi$. It has been shown~\cite{Profumo:2003hq} that the relic abundance enhancement at $\xi\gg1$ scales as $\Delta\Omega\propto \sqrt{\xi}$, and that the maximal quintessential enhancement compatible with BBN bounds is $\Delta\Omega\equiv(\Omega_{\phi}-\Omega_{{\rm no}\ \phi})/\Omega_{{\rm no\ }\phi}\sim 10^6$.

\begin{figure}[!t]
\begin{center}
\mbox{\hspace{-1.5cm}\epsfig{file=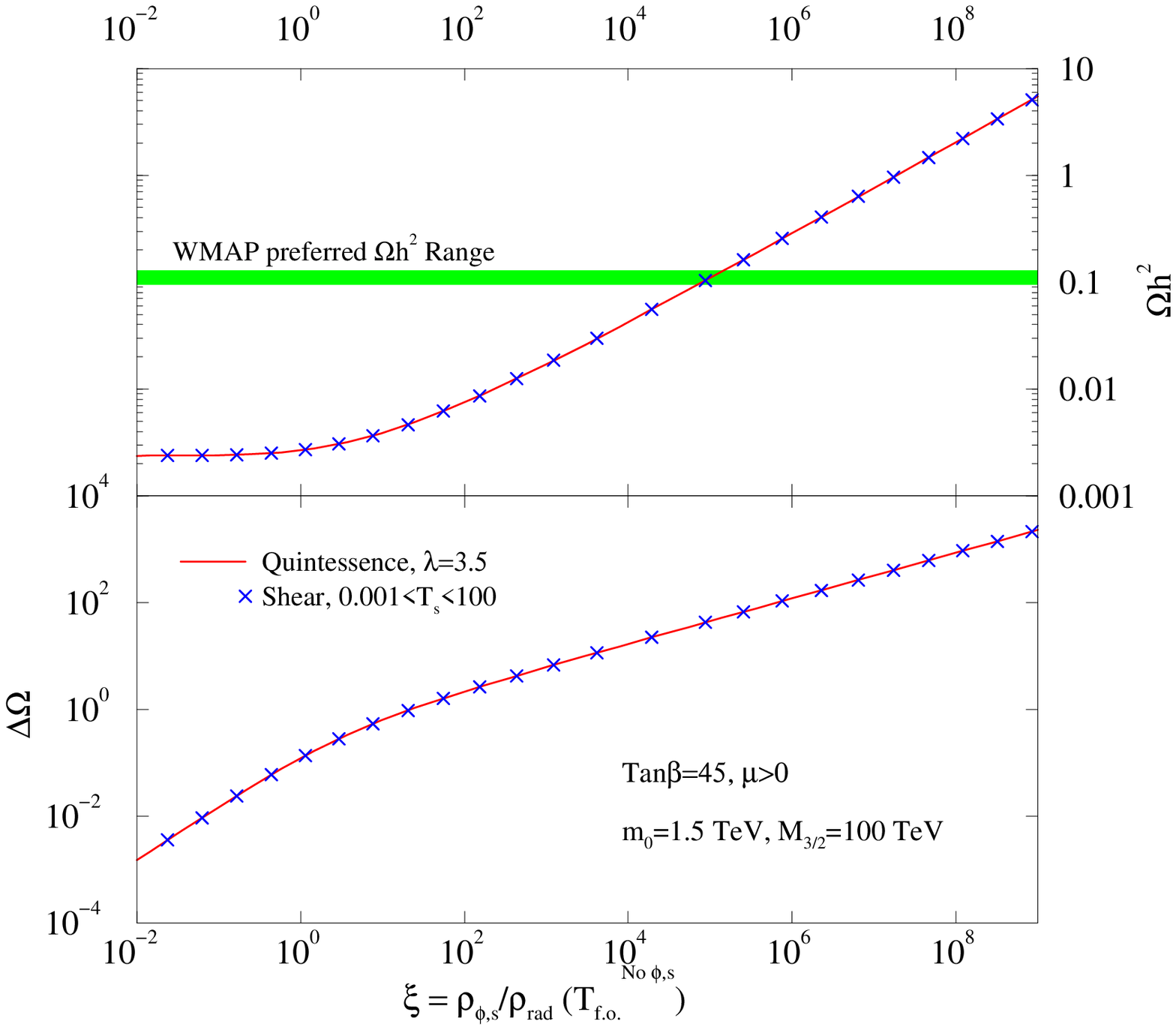,width=9.cm}  \epsfig{file=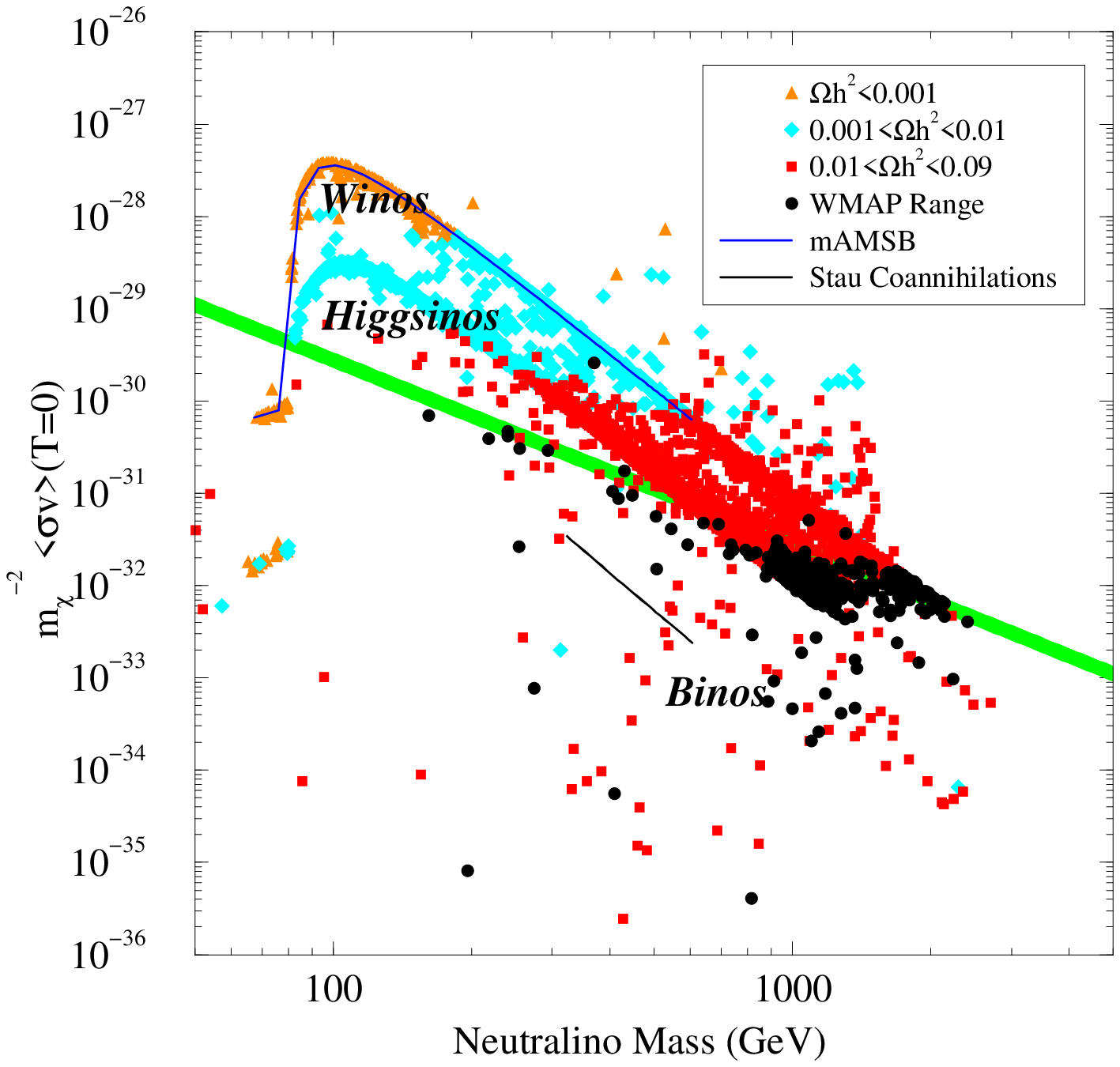,width=8.cm}}\\
{\large (a)\hspace{8.5cm}(b)}
\end{center}
\caption{(a): The relic abundance $\Omega_\chi h^2$ and the relative enhancement $\Delta\Omega_\chi$ for the mAMSB model at $\tan\beta=45$, $\mu>0$, $m_0=1.5\ {\rm TeV}$ and $m_{3/2}=100$ TeV as a function of the energy density ratio of non-standard over standard components at the temperature of neutralino freeze-out ($\xi$). The red lines correspond to quintessential models with an exponent $\lambda=3.5$, while blue crosses represent models with primordial anisotropies of various initial size. (b): A scatter plot of supersymmetric models in the plane ($m_\chi,\langle\sigma_{\rm eff}v\rangle_{T=0}/m^2_\chi$). The various symbols correspond to different relic abundance ranges in the standard $\Lambda$CDM cosmological model. The blue line indicates the same mAMSB slice as in fig.~\ref{fig1} (b), while the black line represents the stau coannihilations strip points at $\tan\beta=50$, $A_0=0$ and $\mu>0$ in mSUGRA, compatible with the WMAP range. The green band corresponds to the {\em naively expected} $\langle\sigma_{\rm eff}v\rangle$ from the rule of thumb $\Omega h^2 \simeq 3\cdot 10^{-27} {\rm cm}^3{\rm s}^{-1}/\langle\sigma_{\rm eff}v\rangle$, for  $\Omega h^2$ in the WMAP preferred range.
\label{fig2}}
\end{figure}

Other cosmological scenarios where $H(T)$ is significantly larger than in the $\Lambda$CDM model at neutralino decoupling range from scalar-tensor~\footnote{In the case of scalar-tensor theories it has been found that re-annihilations of neutralinos may however reduce the expected size of the relic abundance enhancement\cite{Catena:2004ba}} theories~\cite{Kamionkowski:1990ni,Catena:2004ba} to homogeneous but anisotropic cosmologies~\cite{Kamionkowski:1990ni}: in this latter framework, for the simplest case of Bianchi-I type space-times, where the metric reads
\begin{equation}
{\rm d}s^2=-{\rm d}t^2+R^2_1(t)({\rm d}x_1)^2+R^2_2(t)({\rm d}x_2)^2+R^2_3(t)({\rm d}x_3)^2,
\end{equation}
an effective shear energy density, scaling as $\rho_s\sim a^{-6}$, rapidly falls off, but may well dominate at $T_{\rm f.o.}$. The initial size of the shear component, which can be quantified through the temperature $T_s$ at which $\rho_s=\rho_r$, dictates the value of the parameter $\xi$, which again faithfully parameterizes the thermal relic abundance enhancement (fig.~\ref{fig2} (a)).

\section{Low Relic Density Models}

Indirect detection signals from neutralino dark matter annihilations products are searched for in various channels: neutrinos or muons from $\chi\chi$ annihilations in the center of the Sun, of the Earth or in the Galactic center, gamma rays from the Galactic center or from other high-density dark matter regions, and antimatter (positrons, antiprotons, antideuterons) from the Dark Matter Halo. All indirect detection channels linearly depend on the effective neutralino annihilation cross section, $\langle\sigma_{\rm eff} v\rangle$ at $T=0$, and inversely on the square of the neutralino mass $m_\chi$. The first quantity is often, though not always, related to the inverse of the neutralino relic abundance, which is however sensitive to $\langle\sigma_{\rm eff} v\rangle$ at temperatures around neutralino freeze-out. The naive expectation from the $\Omega h^2 \simeq 3\cdot 10^{-27} {\rm cm}^3{\rm s}^{-1}/\langle\sigma_{\rm eff}v\rangle$ relation, for the WMAP range, is shown in fig.~\ref{fig2} (b). The largest deviations (see the black dots in the figure, outside the naive WMAP strip) occur in presence of coannihilations, thresholds or of resonances. Noticeably, the size of the mentioned deviations may be as large as more than five orders of magnitude! In the figure we also indicate the clusters of points corresponding to a given lightest neutralino type (binos, higgsinos, winos). Conspicuous indirect detection signals are generically expected from low relic density models, as one can read out from fig.~\ref{fig2} (b). Models with wino or dominantly higgsino-like lightest neutralinos with $m_\chi\sim\mathcal O (100)$ GeV fall within this category (mAMSB, SUGRA with non-universal gaugino masses). In this respect, not only does the cosmological thermal relic abundance enhancement render low relic density models {\em compatible} with the CDM content of the Universe, but it also implies {\em promising indirect detection signals} at future facilities~\cite{preparation}.

\section*{Acknowledgments}
S.~Profumo wishes to acknowledge financial support from SISSA and from a European Union Grant for Young Scientists.

\end{document}